\documentclass[showkeys,showpacs,superscriptaddress,nofootinbib]{revtex4-2}
\usepackage{amsmath}
\usepackage{amssymb}
\usepackage{graphicx}
\usepackage{float}
\usepackage{color}
\usepackage{slashed}
\begin{document}

\title{Dirac stars in modified $F(R) $ gravity 
}

\author{Zhuldyz Alibekova}
\affiliation{
Department of Theoretical and Nuclear Physics, Al-Farabi Kazakh National University, Almaty 050040, Kazakhstan
}

\author{
Vladimir Dzhunushaliev
}
\email{v.dzhunushaliev@gmail.com}
\affiliation{
Department of Theoretical and Nuclear Physics, Al-Farabi Kazakh National University, Almaty 050040, Kazakhstan
}
\affiliation{
Institute for Experimental and Theoretical Physics, Al-Farabi Kazakh National University, Almaty 050040, Kazakhstan
}
\affiliation{Academician J.~Jeenbaev Institute of Physics of the NAS of the Kyrgyz Republic, 265 a, Chui Street, Bishkek 720071, Kyrgyzstan}

\author{Vladimir Folomeev}
\email{vfolomeev@mail.ru}
\affiliation{Academician J.~Jeenbaev Institute of Physics of the NAS of the Kyrgyz Republic, 265 a, Chui Street, Bishkek 720071, Kyrgyzstan}

\author{Sergei~Odintsov}
\email{odintsov@ice.cat} 
\affiliation{ICREA, Passeig Luis
Companys, 23, 08010 Barcelona, Spain} 
\affiliation{Institute of
Space Sciences (IEEC-CSIC) C. Can Magrans s/n, 08193 Barcelona,
Spain}




\begin{abstract}
We study Dirac mini-stars in modified $R^2$ gravity, where gravity is sourced by both linear and nonlinear spinor fields. Two-dimensional (2D) and three-dimensional (3D) profiles of the mini-star mass are constructed as functions of the spinor frequency (for the 2D profiles) and the central value of the spinor field (for the 3D profiles) in general relativity and modified gravity for various values of the spinor nonlinearity parameter. We show that the modification of gravity alters the properties of these configurations, while increasing the nonlinearity parameter within the framework of $R^2$ gravity significantly changes the profiles of the spinor and metric functions, the effective energy density, and the dependence of the total mass on the oscillation frequency. The most pronounced changes occur at large values of the nonlinearity parameter.
\end{abstract}



\maketitle

\section{Introduction}

One of the topical problems in modern gravitation is the study of compact self-gravitating configurations formed by fundamental fields. Among such objects, particular interest is drawn to systems supported by spinor (fermionic) fields, known as Dirac stars \cite{Herdeiro:2017fhv}.

Within the framework of general relativity (GR), Dirac stars are described by a coupled system of Einstein and Dirac equations in curved spacetime. The spinor field affects the geometry through its energy-momentum tensor, while the geometry, in turn, governs the dynamics of the spinor field. The first regular solutions of this kind were obtained in Refs.~\cite{Finster:1998ws,Finster:1998ux}. Subsequently, families of asymptotically flat solutions were constructed in which the source of gravity is either a linear~\cite{Herdeiro:2017fhv,Herdeiro:2019mbz,Herdeiro:2021jgc} or a nonlinear spinor field~\cite{Krechet:2014nda,Adanhounme:2012cm,Dzhunushaliev:2018jhj,Bronnikov:2019nqa,Dzhunushaliev:2019kiy,Dzhunushaliev:2019uft}.

In turn, from the perspective of cosmological applications, Ref.~\cite{Saha:2016cbu} investigated Friedmann--Robertson--Walker models in which nonlinear spinor fields act as the source of gravity. It was shown that a self-interacting spinor field can effectively mimic various forms of matter, including dust, radiation, quintessence, phantom energy, and Chaplygin gas. In particular, by choosing an appropriate nonlinear term in the spinor-field Lagrangian, one can reproduce a prescribed equation of state, thereby describing both the accelerated expansion of the Universe and more complex evolutionary scenarios, including oscillatory and quintom regimes. Thus, spinor fields can serve as an effective model of dark energy. An important advantage of this approach is that the Bianchi identity is satisfied automatically, considerably simplifying the analysis of multicomponent cosmological models.

On the other hand, Ref.~\cite{Konoplya:2021hsm} investigated traversable wormholes in Einstein--Dirac--Maxwell theory, in which gravity is sourced by two Dirac fields and an electromagnetic field, all minimally coupled. The authors developed an earlier model of a wormhole supported by a spinor field and demonstrated the possibility of asymptotically flat traversable wormhole solutions without the need to introduce phantom matter (see also Ref.~\cite{Dzhunushaliev:2025ntr} for rotating wormholes in Einstein--Dirac--Maxwell theory).
In contrast to previously obtained reflection-symmetric solutions, which exhibited several difficulties, including nonsmooth fields at the throat, the presence of a thin matter shell, and the coexistence of particles and antiparticles without annihilation~\cite{Bolokhov:2021fil,Danielson:2021aor,Konoplya:2021hsm}, smooth configurations asymmetric with respect to the throat were found that are free from these shortcomings. It was shown that the corresponding solutions possess smooth metric and spinor field profiles, are normalizable, and are asymptotically flat. These properties make such configurations more physically plausible and open up the possibility of traversable wormholes supported by ordinary spinor fields within the framework of GR.

Nevertheless, despite its outstanding experimental successes, GR has fundamental limitations. It predicts the existence of singularities, faces difficulties in explaining the observed accelerated expansion of the Universe without introducing additional components, such as dark energy, and is incompatible with the principles of quantum mechanics in the high-energy regime. These issues have motivated extensive research into modified gravity theories  (MGTs), particularly $F(R)$ gravity.

From a cosmological perspective, MGTs provide an alternative approach to explaining the nature of dark energy, making it possible to reproduce the accelerated expansion of the Universe without introducing exotic forms of matter. In particular, $F(R)$-type models can effectively describe both the early Universe and its late-time cosmological evolution through additional geometric degrees of freedom~\cite{Capozziello:2011et,DiValentino:2025,Nojiri:2010wj,Nojiri:2017ncd}.

In the present work, we investigate Dirac stars within the framework of modified $R^2$ gravity, one of the most physically well-motivated alternatives to GR. This model was proposed by Starobinsky as early as 1980~\cite{Starobinsky1980} in the context of inflationary scenario for the early Universe and remains widely used in astrophysics and cosmology to this day. The study of Dirac stars in this theory makes it possible to elucidate the influence of additional geometric degrees of freedom on the structure and physical properties of compact objects.

\section{Statement of the problem and general equations}
\label{prob_statem}

We work within the framework of MGT and investigate compact starlike systems composed of a self-gravitating spinor field. The action for such a system can be written in the following form [we adopt the metric signature $(+,-,-,-)$ and use natural units $c=\hbar=1$]:
\begin{equation}
	S_{\text{tot}} = - \frac{1}{16\pi G}\int d^4 x
		\sqrt{-g} F(R) +S_{\text{sp}} ,
\label{action_tot}        
\end{equation}
where $G$ is Newton's gravitational constant, $F(R)$ is an arbitrary function of the Ricci scalar $R$, and $S_\text{sp}$ denotes the action of the classical spinor field $\psi$, whose Lagrangian is given by
\begin{equation}
	L_{\text{sp}} =	\frac{\imath}{2} \left(
			\bar \psi \gamma^\mu \psi_{; \mu} -
			\bar \psi_{; \mu} \gamma^\mu \psi
		\right) - \mu  \bar \psi \psi - F(S).
\label{lagr_sp}
\end{equation}
Here, $\mu$ denotes the mass of the spinor field, and the semicolon denotes the covariant derivative, defined as
$
\psi_{; \mu} =  [\partial_{ \mu} +1/8\, \omega_{a b \mu}\left( \gamma^a  \gamma^b- \gamma^b  \gamma^a\right)]\psi
$.
The matrices $\gamma^a$ are the Dirac matrices in the standard representation in flat spacetime [see, e.g., Ref.~\cite{Lawrie2002}, Eq.~(7.27)]. The Dirac matrices in curved spacetime, $\gamma^\mu$, are related to the flat-spacetime matrices $\gamma^a$ via the tetrads $e_a^{\phantom{a}\mu}$ as follows: $\gamma^\mu = e_a^{\phantom{a}\mu}\gamma^a$. The quantities $\omega_{ab\mu}$ denote the spin connection [see Ref.~\cite{Lawrie2002}, Eq.~(7.135), for its definition]. The Lagrangian~\eqref{lagr_sp} includes an arbitrary nonlinear term $F(S)$ involving the invariant $S$, which may depend on
$
	\left( \bar\psi \psi \right),
	\left( \bar\psi \gamma^\mu \psi \right)
	\left( \bar\psi \gamma_\mu \psi \right)$, or
	$\left( \bar\psi \gamma^5 \gamma^\mu \psi \right)
	\left( \bar\psi \gamma^5 \gamma_\mu \psi \right)$.
	
Before proceeding to the detailed investigation of the system under consideration, we note that classical spinors are, as is well known, Grassmann numbers. This raises the question of how to define the action and Lagrangian for such a spinor field in a mathematically consistent manner. This issue is discussed in Appendix~\ref{appendix}, where a rigorous mathematical formulation of the derivation of the spinor-field Lagrangian in a well-defined form is presented. To this end, the spinor is decomposed by separating its variables: it is represented as the product of an ordinary complex-valued function and a Grassmann number. To obtain a well-defined action, an additional integration over the Grassmann variables is performed, after which the Lagrangian and the action become, respectively, a real number and a real-valued function.

By varying the action \eqref{action_tot} with respect to the metric and the spinor fields, one obtains the gravitational field equations and the Dirac equation in curved spacetime. We choose the arbitrary function $F(R)$ in the form $F(R) = R + \alpha h(R)$, where $h(R)$ is an arbitrary function of the curvature scalar $R$, and $\alpha$ is a parameter (the limit $\alpha = 0$ recovers GR). It should be noted that the parameter $\alpha$ cannot take arbitrary values and must be constrained by observational data. In the case of quadratic $F(R)$ gravity, two main classes of constraints on $\alpha$ can be distinguished. In the weak-field regime, in particular from observations of binary pulsar systems, the constraint
$|\alpha| \lesssim 5 \times 10^{15}\,\text{cm}^2$
has been obtained~\cite{Naf2010}. In the strong-field regime, which is characteristic of compact astrophysical objects, the corresponding constraint is more stringent:
$|\alpha| \lesssim 10^{10}\,\text{cm}^2$~\cite{Arapoglu2011,Capozziello2016}.

In the present work, to illustrate the influence of modified gravity on the properties of configurations formed by a spinor field, we consider the value of the parameter $\alpha = -10^{13}\,\text{cm}^2$. The choice of a negative sign is not arbitrary: as noted in studies of neutron stars modeling within the framework of $F(R)$ gravity \cite{Astashenok2017,Folomeev:2018ioy}, asymptotically decaying solutions for the metric functions, and consequently for the scalar curvature $R$ and the mass function $m(r)$, can be obtained precisely for $\alpha < 0$ with the metric signature adopted here. Otherwise, the mass function $m(r)$ may exhibit oscillatory behavior, in which case it loses its physical interpretation as the enclosed mass of the configuration.

Using Eqs. \eqref{action_tot} and \eqref{lagr_sp}, we obtain the following system of equations:
\begin{align}
  \left(1+\alpha h_R\right) G_{\mu }^\nu
  -\frac{1}{2}\alpha \left(h - R h_R\right)\delta_{\mu }^\nu
  +\alpha \left(\delta_{\mu }^\nu g^{\rho \sigma} - \delta_{\mu }^\rho   g^{\nu \sigma}\right)
  \left(h_R\right)_{;\rho ;\sigma}
  &= 8\pi G T_{\mu }^\nu,
\label{mdfd_grav_eqn}\\
  \imath \gamma^\mu \psi_{; \mu} - \mu \, \psi -\frac{\partial F}{\partial \bar \psi} &= 0,
\label{dirac_eq}
\end{align}
where $G_{\mu }^\nu\equiv R_{\mu }^\nu - \frac{1}{2} \delta_{\mu }^\nu R$ is the Einstein tensor, $h_R\equiv dh/dR$, and the semicolon denotes covariant differentiation. The right-hand side of Eq. \eqref{mdfd_grav_eqn} contains the energy-momentum tensor of the spinor field, which can be written in the following (symmetrized) form:
\begin{equation}
  T_{\mu}^\nu =\frac{\imath}{4}g^{\nu\rho}\left[
	  \bar\psi \gamma_{\mu} \psi_{;\rho} + \bar\psi\gamma_\rho\psi_{;\mu}
		-\bar\psi_{;\mu}\gamma_{\rho }\psi - \bar\psi_{;\rho}\gamma_\mu\psi
	\right]-\delta_\mu^\nu L_{\text{sp}}.
\label{energy_momentum}
\end{equation}

In the present work, the nonlinear term is chosen in the simplest power-law form
$F(S) = - k(k+1)^{-1}\lambda\left(\bar\psi\psi\right)^{k+1}$, where $k$ and $\lambda$ are free parameters of the model. In what follows, for definiteness, we set $k=1$, which yields
\begin{equation*}
  F(S) = -\frac{\lambda}{2}\left(\bar{\psi}\psi\right)^2.
\end{equation*}
For $\lambda > 0$, the nonlinearity corresponds to an attractive interaction, whereas for $\lambda < 0$, it corresponds to a repulsive one~\cite{Dzhunushaliev:2018jhj}. In the absence of a gravitational field, spinor systems with such a nonlinearity have been studied, for example, in Refs.~\cite{Finkelstein:1951zz,Finkelstein:1956,Soler:1970xp}, where the existence of regular finite-energy solutions was demonstrated. In turn, soliton solutions of the nonlinear Dirac equation in curved spacetime were investigated in Ref.~\cite{Mielke:2017nwt}.

For our purposes, it is convenient to choose the spherically symmetric metric in the form
\begin{equation}
	ds^2 = N(r)\sigma(r)^2 dt^2 - \frac{dr^2}{N(r)} -r^2(d \theta^2 + \sin^2 \theta d \varphi^2),
\label{metric}
\end{equation}
where $N(r)=1-2 G m(r)/(r)$, and the function $m(r)$ corresponds to the current mass of the configuration within the sphere of the radius $r$.

In this paper we work within the framework of $R^2$ gravity, for which
$$
F(R)=R+\alpha h(R)\equiv R+\alpha R^2.
$$

For numerical purposes, it is convenient to treat the scalar curvature $R$ as an independent function, whose governing equation is obtained by taking the trace of the equation~\eqref{mdfd_grav_eqn}:
\begin{equation}
	R''=
	-\left(
		\frac{2}{r}	+\frac{N'}{N}	+\frac{\sigma'}{\sigma}	
	\right)R'
	-\frac{1}{6N\alpha}
	\Biggl\{
		8 \pi G
		\left[\mu\left(g^2-f^2\right)
			+\lambda\left(g^2-f^2\right)^2
		\right]
		+R
	\Biggr\}.
\label{R_eq}
\end{equation}

To describe the spinor field, we employ the following stationary ansatz, which is compatible with the spherically symmetric line element \eqref{metric} (see, e.g., Refs.~\cite{Soler:1970xp,Li:1982gf,Li:1985gf,Herdeiro:2017fhv}):
\begin{equation}
	\psi^{T} =  
	\frac{1}{2} e^{-\imath t \Omega }
	\begin{pmatrix}
	 -\imath e^{\frac{1}{2} \imath (\theta -\varphi )} g
	 & \imath e^{-\frac{1}{2} \imath (\theta +\varphi )} g
	 & -e^{-\frac{1}{2} \imath (\theta +\varphi )} f
	 & e^{\frac{1}{2} \imath (\theta -\varphi )} f \\
	 	 -e^{\frac{1}{2} \imath (\theta +\varphi )} g
	 & -e^{-\frac{1}{2} \imath (\theta -\varphi )} g
	 & -\imath e^{-\frac{1}{2} \imath (\theta -\varphi )} f
	 & -\imath e^{\frac{1}{2} \imath (\theta +\varphi )} f
	\end{pmatrix} ,
\label{spinor}
\end{equation}
where $\Omega$ is the spinor frequency, $f(r)$ and $g(r)$ are two real functions depending only on the radial coordinate $r$. This ansatz for the spinor field ensures that spacetime remains static. Each row of the ansatz describes a spin-$\frac{1}{2}$
 fermion, with the spins of the two fermions oriented in opposite directions. This configuration yields a vanishing total spin, spherical symmetry of the entire system, and a spherically symmetric energy-momentum tensor.
 
 Substituting the metric \eqref{metric} and the ansatz \eqref{spinor}  into the modified gravity equations \eqref{mdfd_grav_eqn} and the Dirac equation  \eqref{dirac_eq}, we obtain a system of ordinary differential equations for the functions $f$, $g$, $\sigma$, and $m$. For numerical integration, it is convenient to introduce dimensionless variables, which reduces the number of parameters in the model under consideration. Since the system possesses a natural length scale set by the spinor mass $\mu$, it is appropriate to render the equations dimensionless using the Compton wavelength $\lambda_c=1/\mu $:
\begin{align}
	x &= \frac{r}{\lambda_c}, \quad
	\bar{\Omega} = \frac{\Omega}{\mu}, \quad
	\left( \bar{f}, \bar{g}\right)  = \sqrt{4\pi}\,\lambda_c^{3/2} \frac{\mu}{M_p} \left( f, g\right) , 
	\nonumber \\
	\bar{m} &= \frac{\mu}{M_p^2} m, \quad
	\Sigma = R \lambda_c^2, \quad
	\bar{\alpha} = \frac{\alpha}{\lambda_c^2}, \quad \bar{\lambda}=\frac{\lambda}{4\pi G},
\label{dimensionless}
\end{align}
where $M_p$ is the Planck mass. 

For notational simplicity, omitting the overbars on dimensionless quantities henceforth, we obtain the following system of equations:
\begin{align}
	f'
	&+ f\left[
		\frac{N'}{4N}
		+\frac{\sigma'}{2\sigma}
		+\frac{1}{x}
		\left(
		1+\frac{1}{\sqrt{N}}
		\right)
	\right]
	+ g \left(
		\frac{1}{\sqrt{N}}
		-\frac{\Omega}{N\sigma}
		+\lambda
		\frac{f^2-g^2}{\sqrt{N}}
	\right)
	=0,
\label{eq:f_eq} \\
	g'
	&+ g\left[
		\frac{N'}{4N}
		+\frac{\sigma'}{2\sigma}
		+\frac{1}{x}
		\left(
		1-\frac{1}{\sqrt{N}}
		\right)
	\right]
	+ f\left(
		\frac{1}{\sqrt{N}}
		+\frac{\Omega}{N\sigma}
		+\lambda
		\frac{f^2-g^2}{\sqrt{N}}
	\right)
	=0,
\label{eq:g_eq} \\
m'
&=
\frac{x^2}{1+2\alpha\Sigma}
\left\{
	\Omega
	\frac{f^2+g^2}
	{\sqrt{N}\sigma}
	+\alpha
	\left[
		-\frac{\Sigma^2}{4}
		+N \left(
			\Sigma''
			+\left(
			\frac{N'}{2N}
			+\frac{2}{x}
			\right)\Sigma'
		\right)
	\right]
	+\frac{\lambda}{2}
	\left(f^2-g^2\right)^2
\right\},
\label{eq:m_eq} \\
\sigma'
&=
\frac{x\sigma}
{1+2\alpha\Sigma}
	\left[
			\Omega
			\frac{f^2+g^2}
			{N^{3/2}\sigma}
			-\frac{fg'-gf'
			}{\sqrt{N}}
		+\alpha
		\left(
			\Sigma''
			-\Sigma'
			\frac{\sigma'}
			{\sigma}
		\right)
	\right],
\label{eq:sigma_eq}
\end{align}
where a prime denotes differentiation with respect to the dimensionless radial coordinate $x$, and the function
\[
N(x)=1-\frac{2m(x)}{x}
\]
retains its original form in terms of the dimensionless variables.

Equations \eqref{eq:m_eq} and \eqref{eq:sigma_eq} correspond to the ($^0_0$) and  $[(^0_0)~-~(^1_1)]$ components of the gravitational field equations~\eqref{mdfd_grav_eqn}, respectively. The system is supplemented by Eq. \eqref{R_eq}  for the scalar curvature $R$, which is similarly cast into dimensionless form using the transformations \eqref{dimensionless}. Thus, we obtain a system of five ordinary differential equations for the five unknown functions $f(x)$, $g(x)$, $\sigma(x)$, $m(x)$, and $\Sigma(x)$.

\section{Numerical solutions}
\label{num_sol}

In this section, we perform a numerical investigation of the system of ordinary differential equations \eqref{R_eq}, \eqref{eq:f_eq}-\eqref{eq:sigma_eq}. We first examine the behavior of the solutions in the vicinity of $r=0$ in order to derive constraints on the possible values of the coefficients in the Taylor expansions of the unknown functions. We then investigate the asymptotic behavior of regular solutions. Finally, we use the resulting solutions to calculate an integral quantity, namely, the mass of a Dirac mini-star.

\subsection{Expansion near the origin}

To ensure the regularity of the solutions at the origin, $x=0$, we need to examine the behavior of the functions $f, g, \sigma, m,$ and $\Sigma$ in the vicinity of this point:
$$
	 f \approx  f_1 x,   g \approx  g_c + \frac{1}{2} g_2 x^2, \Sigma \approx \Sigma_c + \frac{1}{2}\Sigma_2 x^2,  \sigma \approx \sigma_c + \frac{1}{2}\sigma_2 x^2,  m \approx \frac{1}{6}  m_3 x^3 .
$$
This form of the expansions follows from the requirement that the functions be regular at the center.
Substituting these expansions into the system of equations allows us to express the coefficients $f_1, g_2, m_3, \sigma_2,$ and $\Sigma_2$ in terms of the central values $g_c, \sigma_c, \Sigma_c$ and the frequency $\Omega$:
\begin{equation*}
\begin{split}
f_1 &=
-\frac{
g_c \sigma_c
- g_c^3 \lambda \sigma_c
- g_c \Omega
}{
3\sigma_c
}
, \\
\sigma_2 &=
-\frac{
8g_c^2\sigma_c
-4g_c^4\lambda\sigma_c
+\Sigma_c\sigma_c
-24g_c^2\Omega
}{
18\left(1+2\alpha\Sigma_c\right)
}, \\
g_2 &=
-\frac{
-4g_c\sigma_c^2
+8g_c^3\lambda\sigma_c^2
-4g_c^5\lambda^2\sigma_c^2
+g_c\Sigma_c\sigma_c^2
+4g_c\Omega^2
}{
12\sigma_c^2
}, \\
m_3 &=
-\frac{
4g_c^2\sigma_c
-2g_c^4\lambda\sigma_c
+2\Sigma_c\sigma_c
+3\alpha\Sigma_c^2\sigma_c
-12g_c^2\Omega
}{
6\left(1+2\alpha\Sigma_c\right)\sigma_c
}, \\
\Sigma_2 &=
-\frac{
2g_c^2
+2g_c^4\lambda
+\Sigma_c
}{
18\alpha
}.
\end{split}
\end{equation*}

The resulting expansions are used to specify the boundary conditions for the numerical integration of the system of equations.

\subsection{Asymptotic behavior and boundary conditions}

The system of equations describing a self-gravitating Dirac field in $R^2$ gravity is a system of nonlinear ordinary differential equations for the functions $f(x), g(x), \sigma(x), m(x)$, and $\Sigma(x)$. We seek regular, localized, asymptotically flat solutions with finite energy, corresponding to Dirac-star configurations.

At the origin, $x = 0$, the boundary conditions are specified as follows:
\begin{align}
	m(0) = 0, f(0) = 0, g(0) = g_c, 
	\sigma(0) = \sigma_c, \Sigma'(0) = 0.
\label{gran_0}
\end{align}
Here $g_c, \sigma_c$ and the central value of the curvature are free parameters to be determined in the process of finding regular solutions.

To obtain an asymptotically flat solution, we impose the following boundary conditions at infinity (as $x \to \infty$):
\begin{equation}
f(x), g(x) \to 0, \quad 
\sigma(x) \to 1, \quad m(x) \to M, \quad \Sigma(x) \to 0,
\label{gran_infty}
\end{equation}
where $M$ is the total mass of the configuration.

An analysis of the asymptotic behavior of the spinor and metric functions shows that the spinor functions $f(x)$ and $g(x)$, as well as the scalar curvature $\Sigma(x)$, are localized near the center of the configuration and decay exponentially at infinity. In contrast, the metric functions $m(x)$ and $\sigma(x)$ approach constant values asymptotically as $x \to \infty$, with exponentially decaying corrections:
\begin{equation}
	f(x),\, g(x) \sim e^{-\sqrt{1-\Omega^2}\,x}, 
	m(x) \sim M- \frac{A}{\beta} x^2 e^{-\beta x}, \, \sigma(x) \sim 1- \frac{A}{\beta} x e^{-\beta x}, \, \Sigma(x) \sim \frac{\Sigma_\infty}{x} e^{-\frac{x}{\sqrt{6 |\alpha|}}} 	,
\label{BC_infty}
\end{equation}
where $A, \beta$, and $\Sigma_\infty$ are some constants. 

\subsection{Numerical method}

The system of equations \eqref{R_eq} and \eqref{eq:f_eq}-\eqref{eq:sigma_eq} is solved as a boundary-value problem, with boundary conditions imposed at the origin, \eqref{gran_0}, and at infinity, \eqref{gran_infty}. Numerical analysis shows that regular solutions of this system exist only for specific values of the spinor frequency $\Omega$. Therefore, solving the system of equations \eqref{R_eq} and \eqref{eq:f_eq}-\eqref{eq:sigma_eq} reduces to an eigenvalue problem, with the spinor-field frequency $\Omega$ serving as the eigenvalue.

The system of equations \eqref{R_eq} and \eqref{eq:f_eq}-\eqref{eq:sigma_eq} was numerically integrated using the shooting method, with the spinor frequency $\Omega$ as the shooting parameter. The equations were integrated numerically from a point near the origin, $x=\delta\ll 1$, to a finite boundary point $x=x_b$, where the functions $f$, $g$, and $\Sigma$, together with their derivatives, tend to zero.

\subsection{The mass of the configuration}
The mass density of the configuration under consideration is defined as follows:
\begin{equation*}
	\rho_{\mathrm{eff}} = 	\frac{1}{1+2\alpha\Sigma}
	\left\{
		\Omega
		\frac{f^2+g^2}
		{\sqrt{N}\sigma}
        +\frac{\lambda}{2}
		\left(f^2-g^2\right)^2
		+\alpha
		\left[
			-\frac{\Sigma^2}{4}
			+N \left(
				\Sigma''
				+\left(
				\frac{N'}{2N}
				+\frac{2}{x}
				\right)\Sigma'
			\right)
		\right]
	\right\} = \rho_{\mathrm{sp}} + \rho_{\mathrm{MG}} , 
\label{en_dens}
\end{equation*}
where
\begin{align*}
	\rho_{\mathrm{sp}} &=
	\frac{1}{1+2\alpha\Sigma}
	\left[
	\frac{\Omega(f^2+g^2)}{\sqrt{N}\sigma}
	+\frac{\lambda}{2}(f^2-g^2)^2
	\right],
\\
	\rho_{\mathrm{MG}} &=
	\frac{\alpha}{1+2\alpha\Sigma}
	\left[
		-\frac{\Sigma^2}{4}
		+N \left(	\Sigma'' + \left(\frac{N'}{2N}+\frac{2}{x}\right)\Sigma'
		\right)
	\right] .
\end{align*}
Here $\rho_{\mathrm{sp}}$ represents the contribution of the linear ($\lambda = 0$) and nonlinear ($\lambda \neq 0$) spinor fields, taking into account the modification of gravity for $\alpha \neq 0$, whereas $\rho_{\mathrm{MG}}$ describes the additional contribution arising from the quadratic correction $\alpha R^2$ to the gravitational Lagrangian. Note that the energy density of the spinor field, $T^0_{\ 0}$, is determined from Eq.~\eqref{energy_momentum}.

The mass $M$ of a Dirac mini-star in GR ($\alpha = 0$) and in modified gravity ($\alpha \neq 0$) is given by the following integral:
\begin{equation*}
	\bar{M} \equiv \frac{\mu}{M_p^2} M = \int \limits_0^\infty 
	x^2 \rho_{\mathrm{eff}} dx ,
\label{total_mass}
\end{equation*}
where $\bar{M}$ is the dimensionless mass. 
 To verify the accuracy of the numerical calculations, this value was compared with the asymptotic value of the mass function $m(x)$:
\begin{equation*}
	\bar{M} = \lim_{x \to \infty} m(x).
\label{m_asymp}
\end{equation*}

\subsection{Results of numerical calculations
}

\begin{figure}[H]
    \centering
       \begin{minipage}[t]{.49\linewidth}
        \centering
        \includegraphics[width=1.\linewidth]{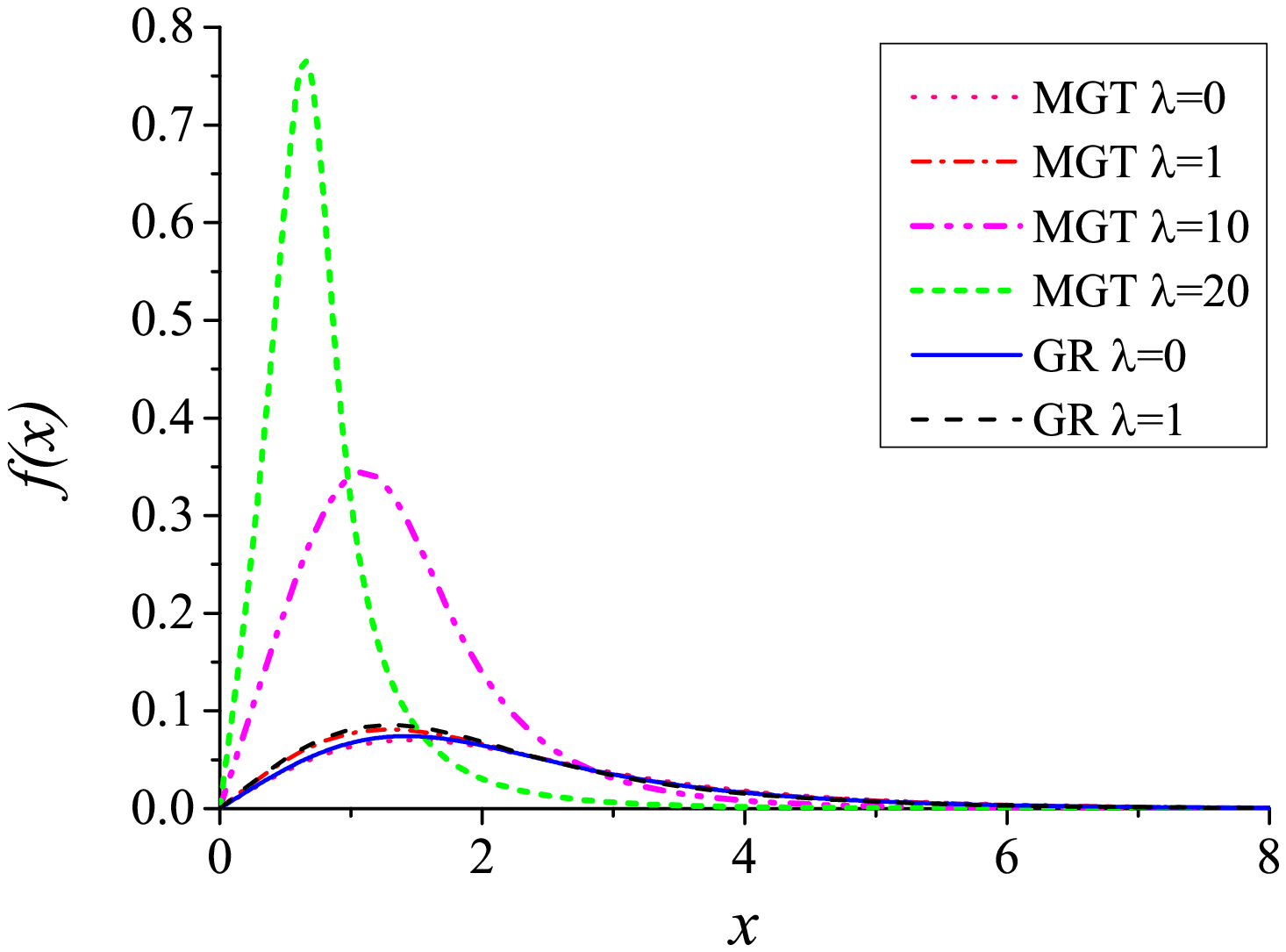}
    \end{minipage}
    \hfill
      \begin{minipage}[t]{.49\linewidth}
        \centering
        \includegraphics[width=1.\linewidth]{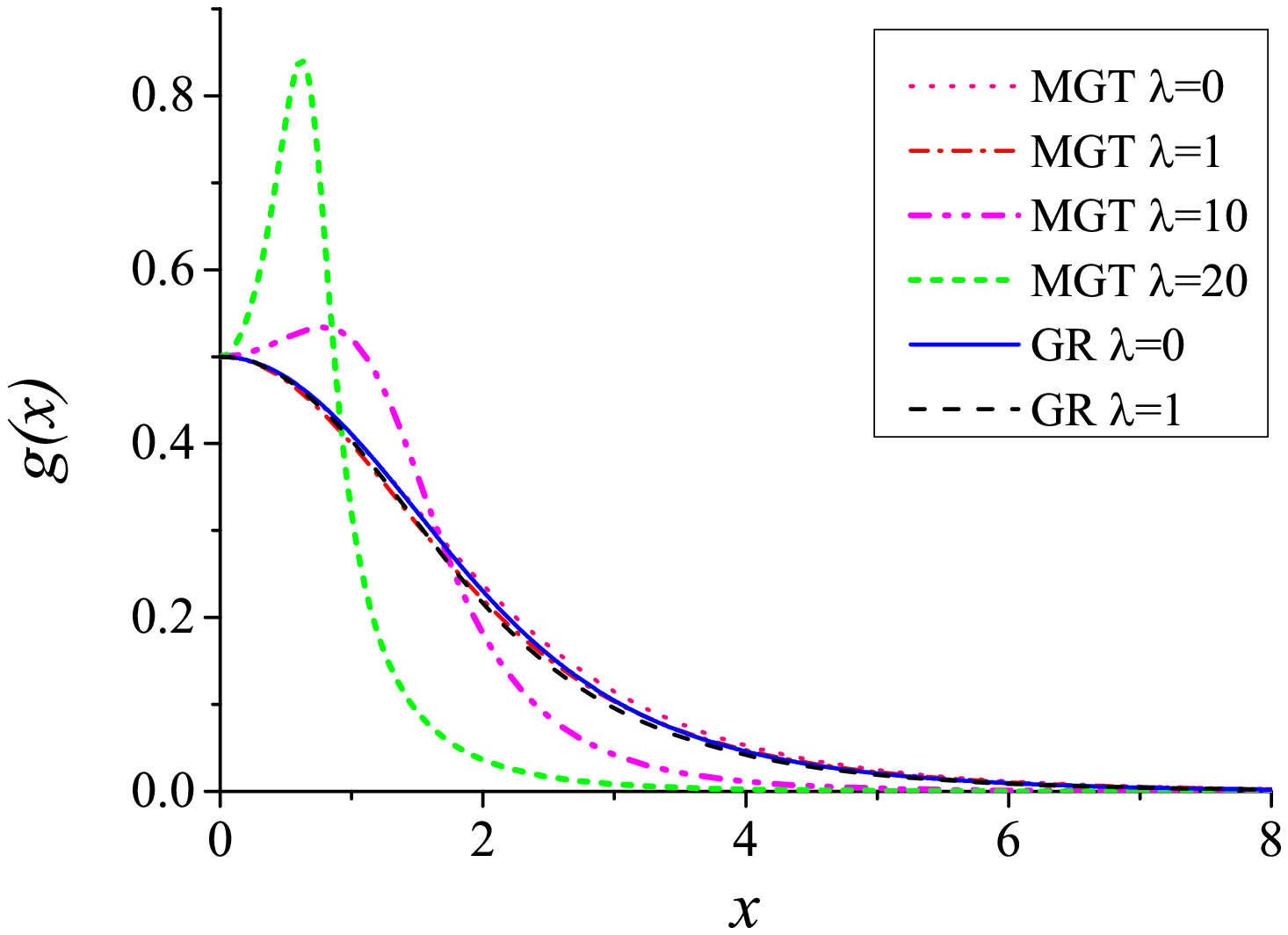}
    \end{minipage}
\vspace{-0.5cm}        
        \caption{Profiles of the spinor functions $f(x)$ and $g(x)$ for different theories of gravity with linear/nonlinear Dirac equations and various values of the nonlinearity parameter $\lambda$.
        } 
        \label{fig_f_g}
\end{figure}

Figure~\ref{fig_f_g} shows representative profiles of the functions $f(x)$ and $g(x)$ for GR and MGT at a fixed value of $g_c = 0.5$ and for different values of the parameter $\lambda$, which characterizes the degree of nonlinearity of the Dirac equation. 
The parameters listed in Table~\ref{tab:params} were used to obtain these profiles.

\begin{table}[H]
\centering
\caption{Parameters used to construct solutions in GR and MGT with linear/nonlinear Dirac equations.
}
\label{tab:params}
\begin{tabular}{l c c c c}
Model & $\lambda$ & $\Omega$ & $\sigma_c$ & $\Sigma_c$ \\
\hline

GR + linear Dirac        & 0  & 0.769587 & 0.509157 & --- \\
GR + nonlinear Dirac     & 1  & 0.777694 & 0.554303 & --- \\

MGT + linear Dirac       & 0  & 0.764015 & 0.514048 & -0.00985844998 \\
MGT + nonlinear Dirac    & 1  & 0.77577 & 0.55795 & -0.00978335 \\
MGT + nonlinear Dirac    & 10 & 0.465734 & 0.463247 & -0.0151279 \\
MGT + nonlinear Dirac    & 20 & 0.474241 & 0.241684 & -0.007895 
\end{tabular}
\end{table}

As shown in Fig.~\ref{fig_f_g}, for $\lambda=0$ and $\lambda=1$, the profiles of the spinor functions $f(x)$ and $g(x)$ remain similar in GR and $F(R)$ gravity. In MGT, increasing the nonlinearity parameter leads to substantial changes in the shape and amplitude of the spinor functions, with the most pronounced effects observed for $\lambda=10$ and, especially, $\lambda=20$. In all cases considered, the functions remain localized and decay exponentially to zero as the radial coordinate increases, in agreement with their asymptotic behavior given by Eq.~\eqref{BC_infty}.

\begin{figure}[h!]
    \centering
    \begin{minipage}{0.495\textwidth}
        \centering
        \includegraphics[width=\textwidth]{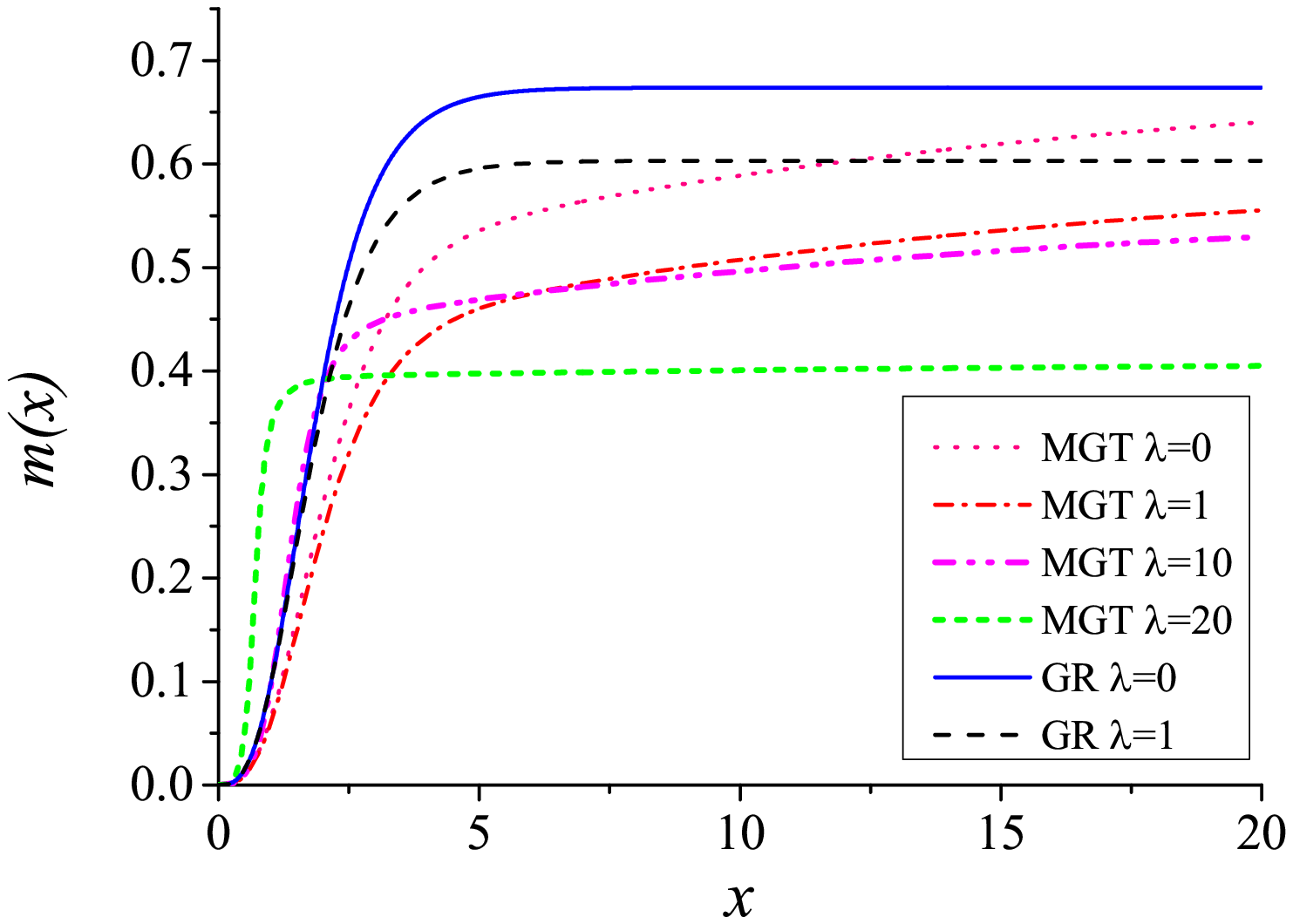}
    \end{minipage}
    \hfill
    \begin{minipage}{0.495\textwidth}
        \centering
        \includegraphics[width=\textwidth]{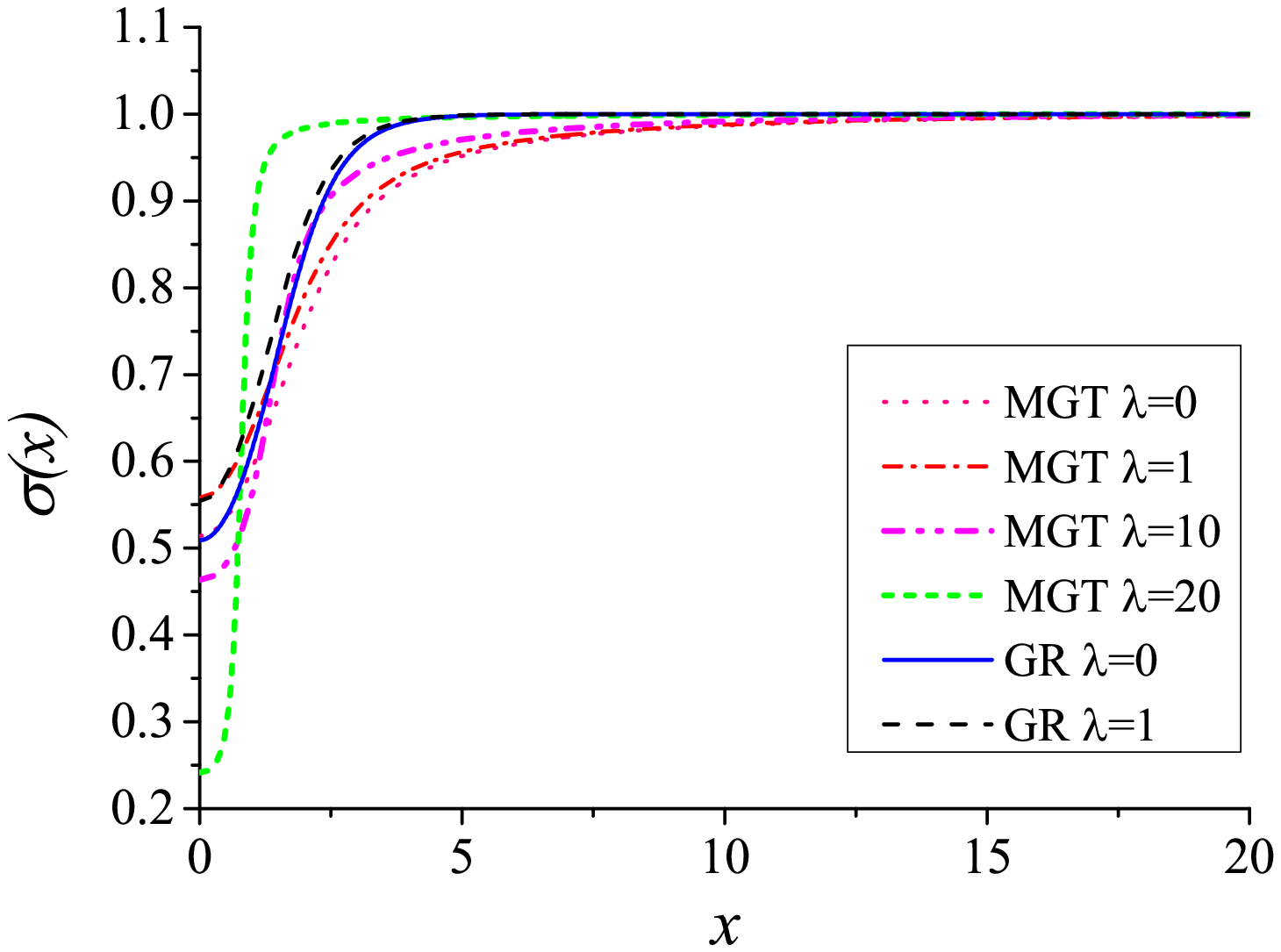}
    \end{minipage}
\vspace{-0.5cm}
\caption{Profiles of the dimensionless mass function $m(x)$ (left panel) and the metric function $\sigma(x)$ (right panel) for different theories of gravity with linear and nonlinear Dirac equations and various values of the nonlinearity parameter $\lambda$.
}
        \label{m_sigma}
\end{figure}

Figure~\ref{m_sigma} shows the profiles of the mass function $m(x)$ and the metric function $\sigma(x)$. For $\lambda=0$ and $\lambda=1$, the corresponding profiles differ between GR and $F(R)$ gravity, with the asymptotic values of $m(x)$, which determine the total mass of the configuration, being smaller in MGT than in GR. Within MGT, increasing the nonlinearity parameter also leads to noticeable changes in both metric functions, particularly for $\lambda=10$ and $\lambda=20$. Nevertheless, for all solutions considered, $\sigma(x)\to 1$, while $m(x)$ approaches a finite asymptotic value as $x\to\infty$.

\begin{figure}[t!]
\begin{minipage}{0.495\textwidth}
    \centering
    \includegraphics[width=\textwidth]{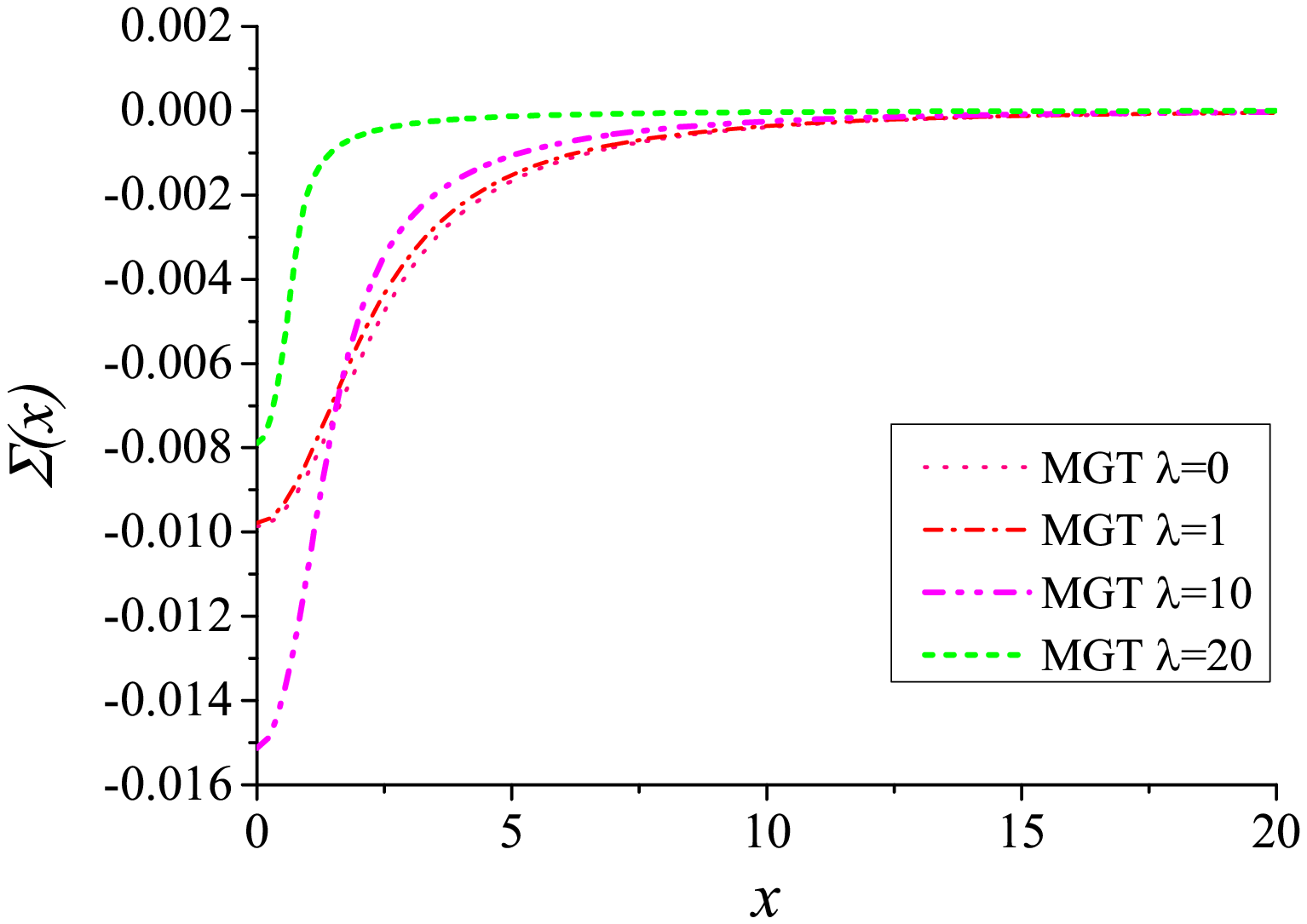}
\vspace{-0.5cm}
    \caption{Profiles of the dimensionless scalar curvature $\Sigma(x)$ for different values of the nonlinearity parameter, $\lambda=0,1,10,20$. The solution parameters are listed in Table~\ref{tab:params}.
    }
    \label{fig_big_sigma}
\end{minipage}
\hfill
\begin{minipage}{0.495\textwidth}
    \centering
    \includegraphics[width=\textwidth]{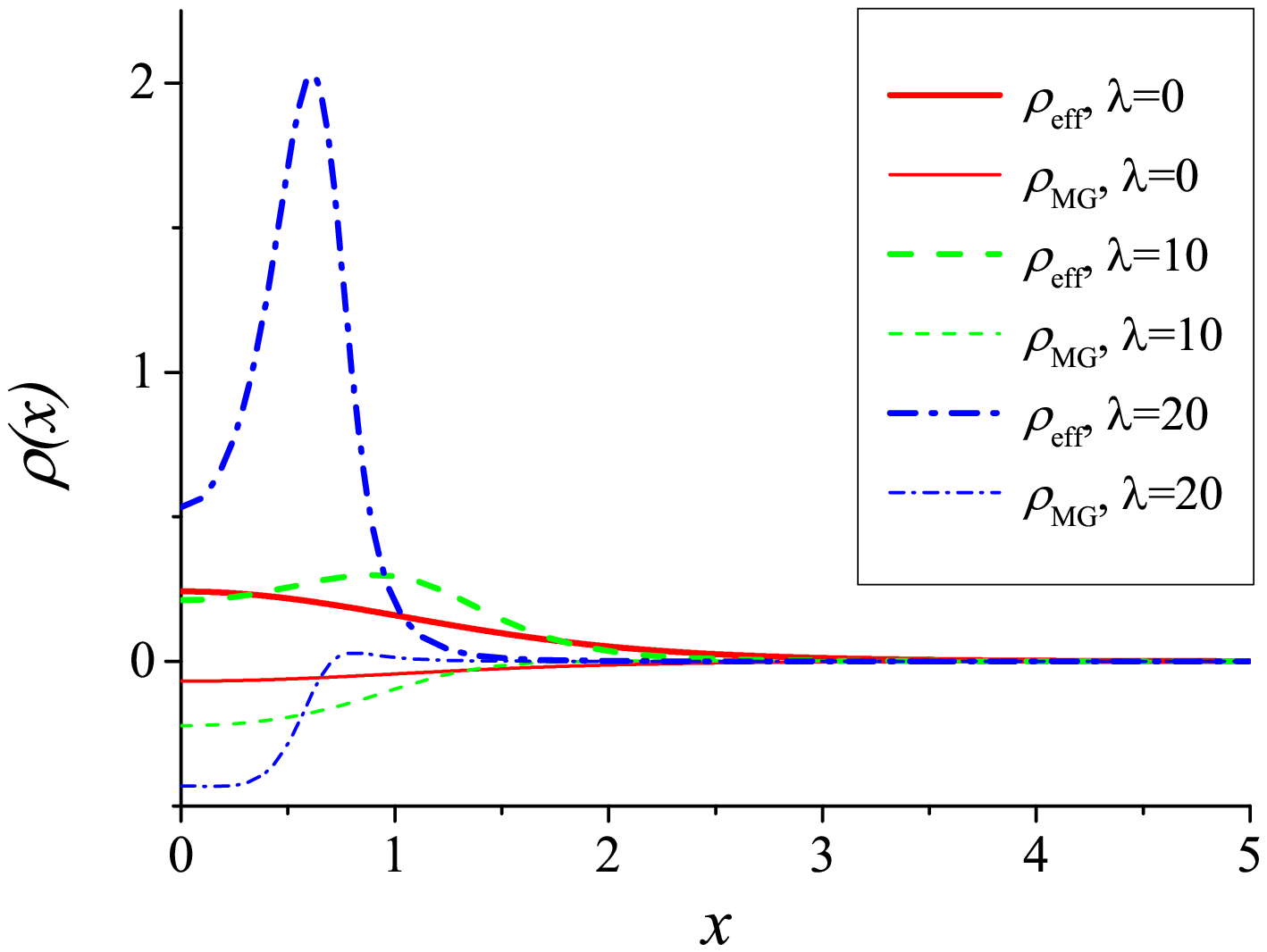}
\vspace{-0.5cm}
    \caption{Profiles of the effective energy density $\rho_{\mathrm{eff}}(x)$ and the geometric contribution $\rho_{\mathrm{MG}}(x)$ for different values of the nonlinearity parameter, $\lambda=0,10,20$. The solution parameters are listed in Table~\ref{tab:params}.
    }
    \label{fig_rho_eff_mg}
\end{minipage}
\end{figure}

The numerical calculations yield the scalar curvature profile $\Sigma$, shown in Fig.~\ref{fig_big_sigma}. As expected, the scalar curvature attains its maximum absolute value at the center of the configuration and tends to zero at infinity. This result is consistent with the required asymptotic behavior given by Eq.~\eqref{BC_infty} and confirms the asymptotic flatness of the obtained solutions.

Figure~\ref{fig_rho_eff_mg} shows the radial distributions of the effective energy density $\rho_{\mathrm{eff}}$ and the geometric contribution $\rho_{\mathrm{MG}}$ associated with the modification of gravity. As the nonlinearity parameter increases, the distribution profiles change significantly: a pronounced maximum develops at a finite distance from the center of the configuration for $\lambda=10$ and, especially, for $\lambda=20$. At large values of $x$, both the effective energy density and the modified-gravity contribution tend to zero, in agreement with the asymptotic behavior of the obtained solutions.

\begin{figure}[h!]
    \centering
    \includegraphics[width=0.65\textwidth]{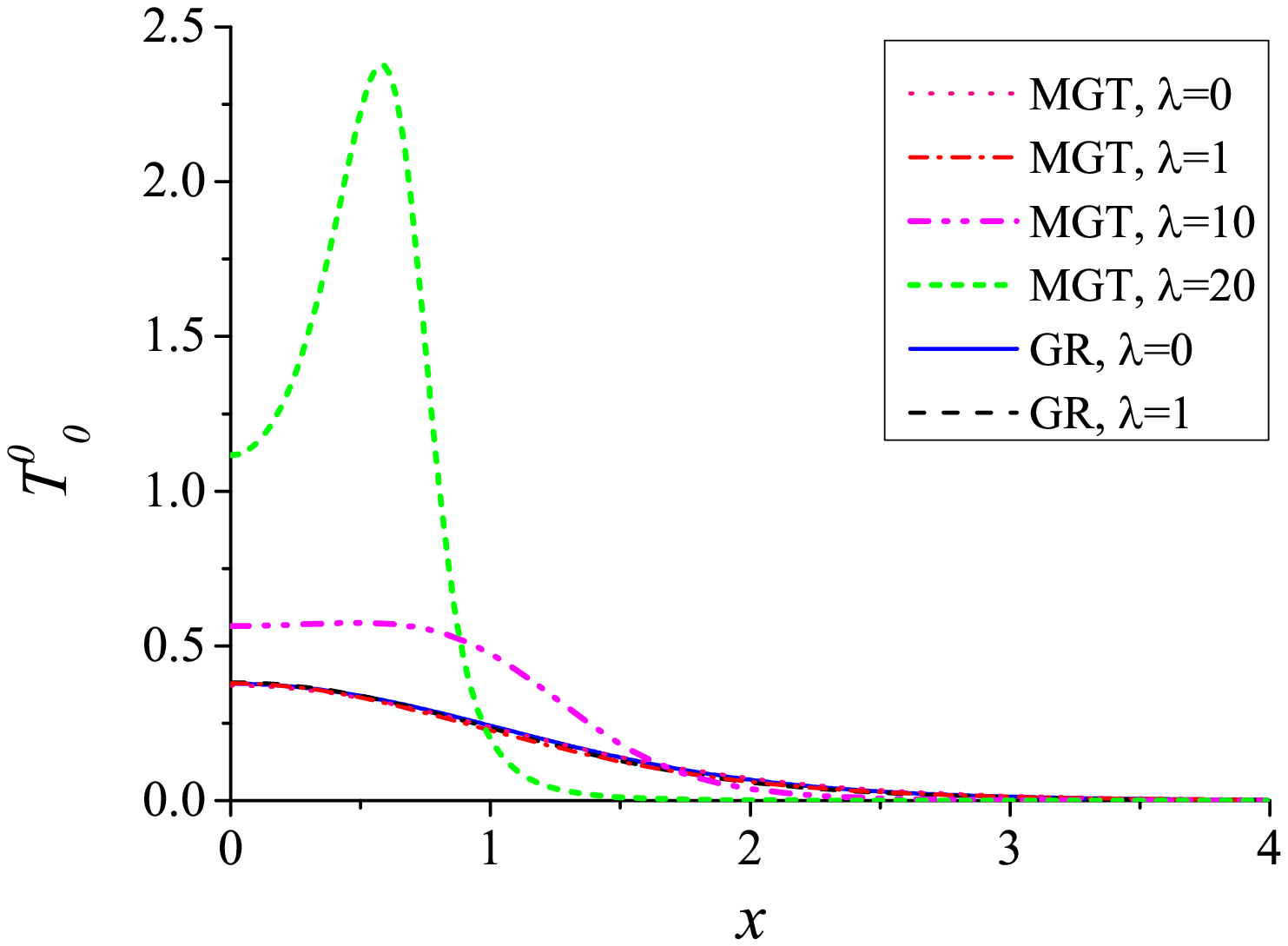}
    \vspace{-0.5cm}
    \caption{Profiles of the spinor-field energy density $T^0_{\ 0}(x)$ in GR and MGT for values of the nonlinearity parameter $\lambda=0,1,10,20$. The solution parameters are listed in Table~\ref{tab:params}.
    }
    \label{fig_t00}
\end{figure}

Figure~\ref{fig_t00} shows the radial distributions of the time component of the energy-momentum tensor~\eqref{energy_momentum}, $T^0_{\ 0}$, of the spinor field for different values of the nonlinearity parameter $\lambda$.

\begin{figure}[t!]
\begin{minipage}{0.495\textwidth}
\centering
\includegraphics[width=\textwidth]{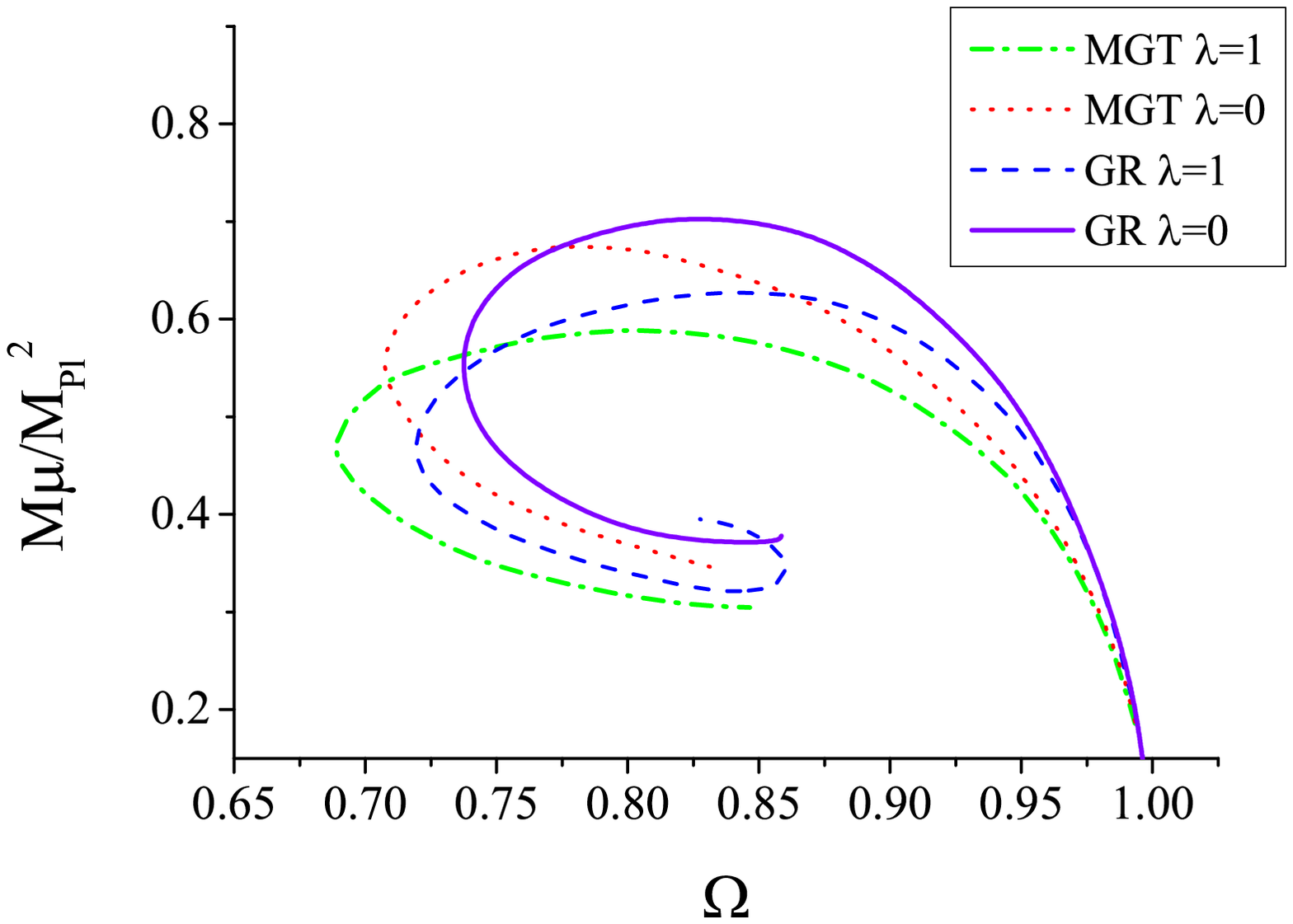}
\vspace{-0.5cm}
\caption{Dependence of the total mass of the configuration on the spinor-field oscillation frequency in GR and MGT for values of the nonlinearity parameter $\lambda=0,1$.
}
\label{mgt_gr}
\end{minipage}
\hfill
\begin{minipage}{0.495\textwidth}
	\centering
\includegraphics[width=\textwidth]{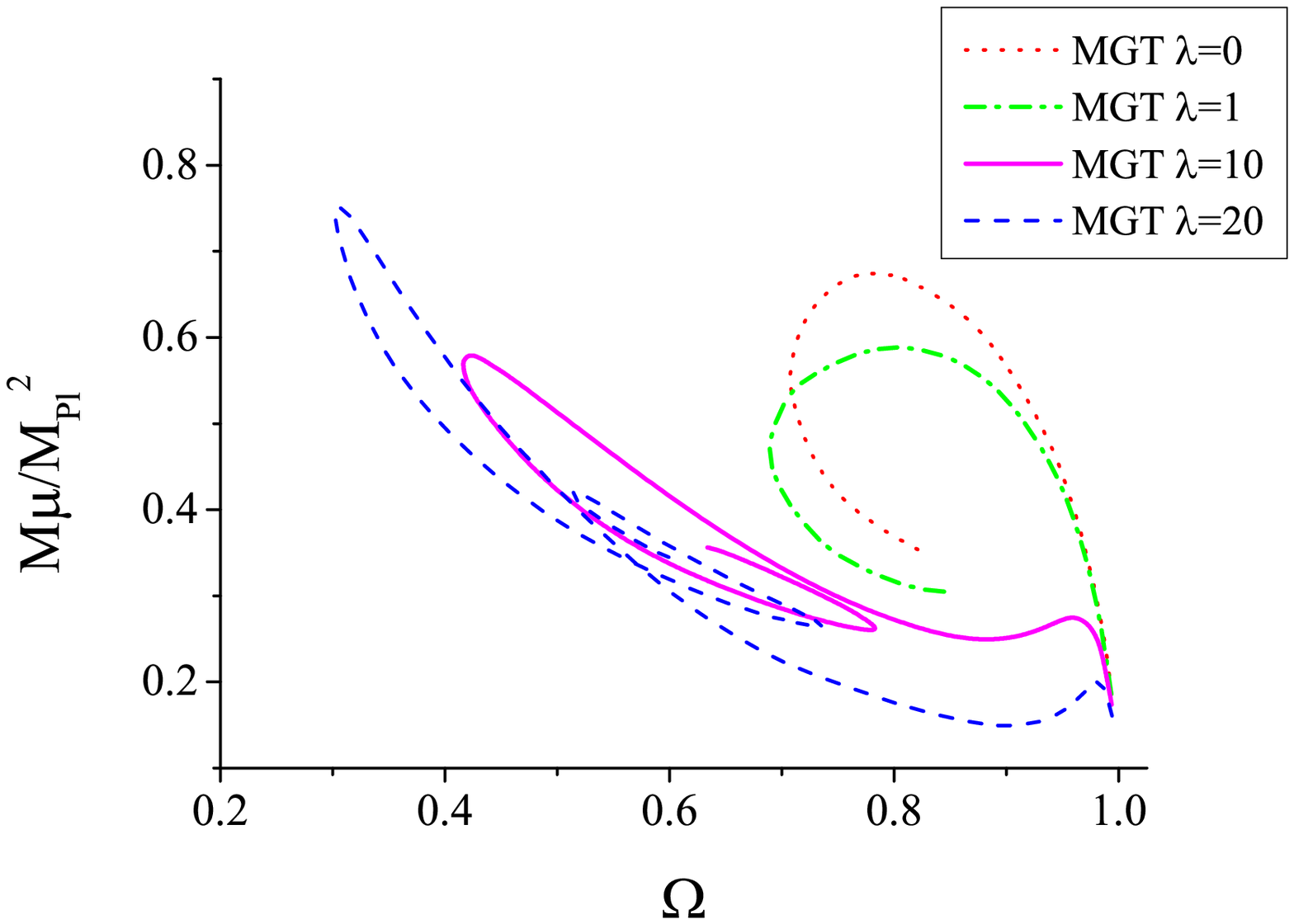}
\vspace{-0.5cm}
\caption{Dependence of the total mass of the configuration on the spinor-field oscillation frequency in MGT for different values of the nonlinearity parameter, $\lambda=0,1,10,20$.
}
\label{mgt_plot}
\end{minipage}
\end{figure}

One of the most important characteristics of a Dirac star is the dependence of its total mass $M$ on the oscillation frequency $\Omega$ of the spinor field. Figure~\ref{mgt_gr} compares the mass-frequency relations $\bar{M}(\Omega)$ in GR and MGT for the linear case, $\lambda=0$, and the nonlinear case, $\lambda=1$. This representation allows a direct comparison of the effects of modifying the theory of gravity on the mass properties of the configurations at identical values of the nonlinearity parameter.

To investigate the influence of the nonlinear spinor field in greater detail, Fig.~\ref{mgt_plot} shows the mass-frequency relations $\bar{M}(\Omega)$ in MGT for values of the nonlinearity parameter $\lambda=0,1,10,20$. It can be seen that varying $\lambda$ changes the shape of the corresponding solution branches, with the most pronounced differences occurring at larger values of the nonlinearity parameter, $\lambda=10$ and $\lambda=20$.

\begin{figure}[H]
    \centering
    \includegraphics[width=0.6\textwidth]{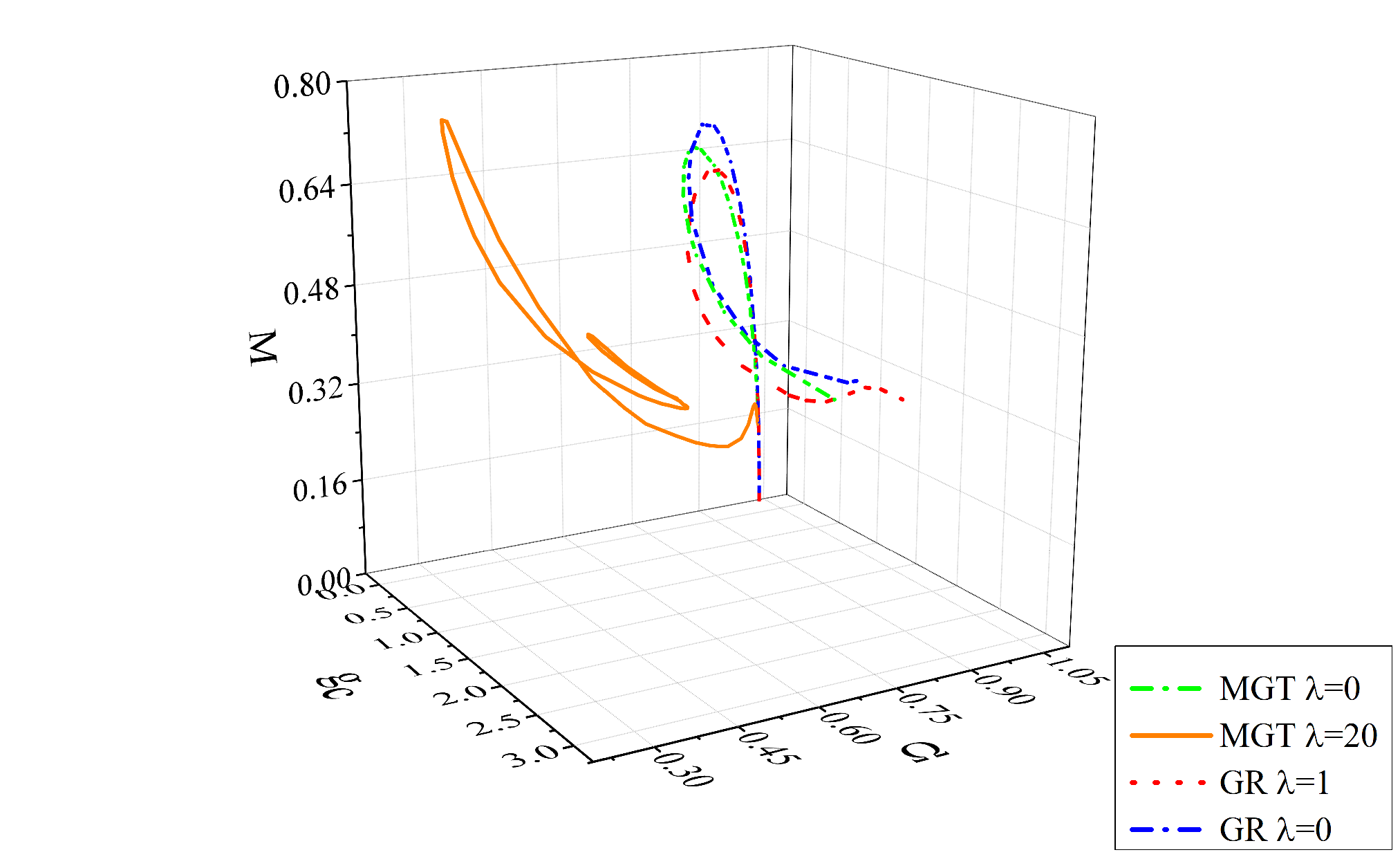}
    \caption{Three-dimensional dependence of the dimensionless total mass $\bar{M}$ on the oscillation frequency $\Omega$ and the central value of the spinor function $g_c$.
    }
    \label{3D_plot}
\end{figure}

In the two-dimensional plots of $\bar{M}(\Omega)$ presented above, certain segments of the solution branches may appear to overlap. To provide a more comprehensive analysis of the solution structure, we have additionally constructed the three-dimensional plot shown in Fig.~\ref{3D_plot}, where the dimensionless mass $\bar{M}$ is considered as a function of the oscillation frequency $\Omega$ and the central value of the spinor function, $g_c=g(0)$. The plot includes GR solutions for $\lambda=0$ and $\lambda=1$, as well as MGT solutions for $\lambda=0$ and $\lambda=20$. This representation makes it possible to trace the positions of the corresponding branches in the parameter space $(\Omega,g_c,\bar{M})$ and demonstrate that the apparent overlap of certain segments in the two-dimensional plots arises from their projection onto the $(\Omega,\bar{M})$ plane.

\section{Conclusion}
\label{concl}

Thus, we have investigated spherically symmetric solutions in modified $R^2$ gravity, with a classical spinor field as the source, and compared the results with those obtained in GR coupled to a linear/nonlinear Dirac equation. Such configurations can be referred to as Dirac mini-stars, following the terminology proposed in Ref.~\cite{Herdeiro:2017fhv}.

The following results have been obtained:
\begin{enumerate}
\item \textit{Total mass}: for the considered cases $\lambda=0$ and $\lambda=1$, at the same values of the frequency $\Omega$, the total mass of the configuration in $F(R)$ gravity is lower than the corresponding mass in GR. Within the MGT framework, increasing the nonlinearity parameter significantly alters the shape of the $\bar M(\Omega)$ dependence, with the most pronounced changes observed, as expected, for $\lambda=10$ and $\lambda=20$.
    
\item \textit{Spinor functions}: for small values of the nonlinearity parameter, the differences between the profiles of $f(x)$ and $g(x)$ in GR and $F(R)$ gravity are relatively small. As $\lambda$ increases, the effects of nonlinear self-interaction become more pronounced, leading to noticeable changes in the shape and amplitude of the spinor functions, particularly for $\lambda=10$ and $\lambda=20$. At large distances, the functions decay exponentially in accordance with the asymptotic behavior $f(x),g(x)\sim e^{-x\sqrt{1-\Omega^{2}}}$.

\item \textit{Metric functions}: the profiles of the mass function $m(x)$ and the metric function $\sigma(x)$ depend on both the gravitational theory under consideration and the nonlinearity parameter. For $\lambda=0$ and $\lambda=1$, differences between the solutions in GR and $F(R)$ gravity are observed, while within the MGT framework, increasing $\lambda$ leads to noticeable changes in both functions.

\item \textit{Scalar curvature}: in GR, the scalar curvature $R$ is algebraically related to the trace of the energy-momentum tensor $T$, whereas in $F(R)$ gravity, the scalar curvature $R$ is related not only to $T$ but also to additional terms arising from the modification of gravity.

\item \textit{Effects of modified gravity and nonlinearity}: the transition from GR to $R^2$ gravity leads to changes in the structure of the solutions, which are reflected in the behavior of the metric functions and the dependence of the total mass on the configuration parameters. Within the framework of $R^2$ gravity, increasing the nonlinearity parameter significantly alters the profiles of the spinor and metric functions, the effective energy density, and the dependence of the total mass on the oscillation frequency of the spinor field, with the most pronounced changes observed for $\lambda=10$ and $\lambda=20$.
\end{enumerate}


\appendix

\section{Lagrangian for classical spinors with Grassmann numbers}
\label{appendix}

The possibility of using a classical spinor field in gravity has not yet been sufficiently explored in the literature. Therefore, in this section, following Ref.~\cite{Dzhunushaliev:2025ngw}, we provide a justification for this possibility.

To begin with, let us recall the definition of a spinor that can be found in standard textbooks, see, e.g., Ref.~\cite{Shifman}. A Weyl spinor is a two complex component object 
$\chi = \begin{pmatrix}
	\chi_1 \\
	\chi_2 
\end{pmatrix} \in \mathbf{C}^2$ transforming under an element 
$
	\mathcal{M} \in \mathrm{Sl}(2, \mathbf{C})
$ as 
$$
	\chi_\alpha \rightarrow \chi^\prime_\alpha = \mathcal{M}_\alpha^{\phantom{\alpha} \beta} \chi_\beta , 
$$
with $\alpha, \beta$ labeling the components. A two-component object $\bar{\chi}$ belonging to the complex conjugate representation and transforming as
$$
	\bar{\chi}_{\dot{\alpha}} \rightarrow \bar{\chi}^\prime_{\dot{\alpha}} = 
	\left(\mathcal{M}^*\right)_{\dot{\alpha}}^{\phantom{\alpha} \dot{\beta}} 
	\bar{\chi}_{\dot{\beta}}
$$
is called a dotted spinor. 

The scalar products $\chi \eta$ and $\bar{\chi} \bar{\eta}$ are defined as
\begin{equation}
	\chi \eta := \chi^{\alpha} \eta_{\alpha} = \eta \chi, \quad 
	\bar{\chi} \bar{\eta} := \bar{\chi}_{\dot{\alpha}} \bar{\eta}^{\dot{\alpha}} 
	= \bar{\eta} \bar{\chi} . 
\label{scalar_prod}
\end{equation}
For the scalar product \eqref{scalar_prod} to be symmetric, we find 
\begin{equation}
	\chi_\alpha \eta_\beta = - \eta_\beta \chi_\alpha , 
\label{Grassman_spinor}
\end{equation}
i.e., spinors must anticommute; this means that the components of the Weyl spinor are Grassmann numbers.

A Dirac spinor $\Psi$ is defined to be the direct sum of two Weyl spinors $\chi$ and $\bar{\eta}$ of opposite chirality 
\begin{equation}
	\Psi := 
	 \begin{pmatrix}
		\chi_\alpha \\
		\bar{\eta}^{\dot{\alpha}}  
	\end{pmatrix}, \quad 
	\bar{\Psi} =  
	 \begin{pmatrix}
		\eta^\alpha , \bar{\chi}_{\dot{\alpha}}  
	\end{pmatrix} . 
\label{Dirac_spinor}
\end{equation}
For a correct definition of the Lagrangian of a spinor field, let us consider the expression
\begin{equation}
	L_D = \imath \bar{\Psi} \gamma^{\mu} D_\mu \Psi - m \bar{\Psi} \Psi 
	=\imath \bar{\Psi} 	 \begin{pmatrix}
			0																&	\sigma^\mu \\
			\bar{\sigma}^\mu	& 0
		\end{pmatrix} D_\mu \Psi - m \bar{\Psi} \Psi 
	= \imath \left[ 
		\bar{\chi}_{\dot{\beta}} \left( \bar{\sigma}\right) ^{\dot{\beta} \alpha} D_\mu \chi_{\alpha} 
		+ \eta^{\alpha} \left( \sigma^\mu \right)_{\alpha \dot{\beta}} D_\mu {\bar{\eta}}^{\dot{\beta}} 
	\right] 
	- m \left( \eta^{\alpha} \chi_{\alpha} + \bar{\chi}_{\dot{\alpha}} \bar{\eta}^{\dot{\alpha}}\right) , 
\label{Dirac_Lagr}
\end{equation}
where 
$
	\sigma^\mu := \left(\mathbb{I}, \sigma^k\right)$ and $\bar\sigma^\mu := \left(\mathbb{I}, -\sigma^k\right)
$, 
and $\chi_\alpha$ and $\bar{\eta}^{\dot{\alpha}}$ are Grassmann numbers. 

In Refs.~\cite{Shifman} (Sec. 7.1) and~\cite{Raja} (Sec. 9.1), it is suggested that the separation of variables is carried out such that one can explicitly extract Grassmann numbers,
$$
	\Psi_\alpha = \sum_{i} \tilde{\Psi}_{i} \theta_{i \alpha} ,
$$
where $\alpha$ is the spinorial index and the coeﬃcients $\theta_{i \alpha}$ are Grassmann numbers.

For our purposes and in order to take into account the fact that spinors \eqref{Dirac_spinor} are Grassmann numbers \eqref{Grassman_spinor}, we choose another decomposition of spinor components into ordinary functions and Grassmann numbers in the following form:
$$
	\Psi := 
	 \begin{pmatrix}
		\chi_\alpha \\
		\bar{\eta}^{\dot{\alpha}}  
	\end{pmatrix} \theta, \quad 
	\bar{\Psi} =  
	 \begin{pmatrix}
		\eta^\alpha , \bar{\chi}_{\dot{\alpha}}  
	\end{pmatrix} \theta^*  ,
$$
where $\theta$ and $\theta^*$ are independent Grassmann numbers, while $\chi_\alpha$ and $\bar{\eta}^{\dot{\alpha}}$ are now complex functions. For such a separation of variables into complex and  Grassmann numbers, the Lagrangian \eqref{Dirac_Lagr} takes the form
$$
	\tilde{L}_D = \left\lbrace \imath 
	\left[ 
		\bar{\chi}_{\dot{\beta}} \left( \bar{\sigma}\right) ^{\dot{\beta} \alpha} D_\mu \chi_{\alpha} 
		+ \eta^{\alpha} \left( \sigma^\mu \right)_{\alpha \dot{\beta}} D_\mu {\bar{\eta}}^{\dot{\beta}} 
	\right] 
	- m \left( \eta^{\alpha} \chi_{\alpha} + \bar{\chi}_{\dot{\alpha}} \bar{\eta}^{\dot{\alpha}}\right) 
	\right\rbrace 
	\theta^* \theta .
$$
Then, introducing integration over the Grassmann variables according to the general rules
$$
	\int 1 d \theta = 0, \quad \int \theta d \theta = 1, \quad 
	\int d \theta d \theta^* \left( \theta^* \theta \right) = 1 , 
$$
one can write out the Lagrangian for a classical spinor field  using integration over the Grassmann variables:
\begin{equation}
	\mathcal{L}_D = \int \tilde{L}_D d \theta d \theta^* 
	= \imath \left[ \bar{\chi}_{\dot{\beta}} \left( \bar{\sigma}\right) ^{\dot{\beta} \alpha} D_\mu \chi_{\alpha} 
		+ \eta^{\alpha} \left( \sigma^\mu \right)_{\alpha \dot{\beta}} D_\mu {\bar{\eta}}^{\dot{\beta}} 
	\right] 
	- m \left( \eta^{\alpha} \chi_{\alpha} + \bar{\chi}_{\dot{\alpha}} \bar{\eta}^{\dot{\alpha}}\right) 
	+ (h. c.) .
\label{Lagrangain_spinor_classic}
\end{equation}
\emph{This gives us a strict, mathematically correct definition of the Lagrangian for a classical spinor field, enabling one to define correctly the current and the energy-momentum tensor for the classical spinor field in terms of ordinary $c$-numbers. } Notice that the integration over the Grassmann variables in \eqref{Lagrangain_spinor_classic} is significantly simpler than that in a path integral when fermion fields are present (see, e.g., Ref.~\cite{Raja}). In this connection also note that, as pointed out in Ref.~\cite{Raja}: ``In the study of the non-second-quantised Dirac equation, standard textbooks usually treat the `Dirac wavefunction' as a $c$-number spinor field, without bothering about Grassmann fields.''

Now it is necessary to address the question in what sense one should understand the limiting transition $\hbar \rightarrow 0 $ from a quantum spinor field to a classical one. As we know, the anticommutation relations for operators of a fermion field are of the form
$$
	\left\lbrace \hat{\Psi}_\alpha \left( x^\mu\right) , \hat{\Psi}_\beta \left(y^\mu\right) 
	\right\rbrace = \frac{1}{\imath} S_{\alpha \beta} \left(x^\mu - y^\mu\right) 
	= \left( 
		\imath \slashed{\nabla}  - m
	\right)_{\alpha \beta} 
	\int \frac{d^4 k}{\left(2 \pi \right)^3 } \epsilon\left( k_0\right) 
	\delta \left(k^2 - m^2 \right) 
	e^{- \imath k \cdot \left(x - y\right) } ,
$$
where $S_{\alpha \beta}$ is a fermion propagator and $\epsilon\left( k_0\right) $ is a step function ($\epsilon=1$ for $k_0>0$ and $\epsilon=-1$ for $k_0<0$).
This expression is written in natural units; if we restore the constants $\hbar$ and $c$,  Planck's constant will appear in the Dirac delta function as
$\delta \left(\hbar^2 k^2 - m^2 c^2\right) $. When $\hbar \rightarrow 0$, we have
$
	\delta \left(\hbar^2 k^2 - m^2 c^2\right) \rightarrow \delta \left(- m^2 c^2\right) = 0 
$. This means that in the limit  $\hbar \rightarrow 0$ the anticommutator 
$
\left\lbrace \hat{\Psi}_\alpha \left( x^\mu\right) , \hat{\Psi}_\beta \left(y^\mu\right) \right\rbrace = 0
$, and this in turn means that the components of the spinor $\Psi$ become Grassmann numbers. 

Thus, let us emphasise the following results:
\begin{itemize}
\item It is shown that in order to take into account the fact that the components of a spinor are Grassmann numbers, one can represent the spinor as the product of ordinary functions and Grassmann numbers. 
\item We gave a strict, mathematically correct definition of the Lagrangian of a spinor field arising after integration over Grassmann variables. Such approach enables one to define correctly the current and the energy-momentum tensor for the classical spinor field.
\item It is shown that in the limit $\hbar \rightarrow 0$ an anticommuting quantum spinor field is transformed into a classical spinor field whose components are Grassmann numbers.
\end{itemize}

\end{document}